\documentclass[reprint,showpacs,amsmath,amssymb,aps,pre]{revtex4-2}
\usepackage{dcolumn}
\usepackage{bm}
\usepackage[dvipdfmx]{graphicx}
\usepackage[dvipdfmx]{hyperref}
\usepackage[mathlines]{lineno}
\usepackage{physics}

\begin{document}

\title{Real-space formulation of the Chern invariant and topological phases in a disordered Chern insulator}
\author{Kiminori Hattori and Shinji Nakata}
\affiliation{Department of Systems Innovation, Graduate School of Engineering Science, The University of Osaka, Toyonaka, Osaka 560-8531, Japan}
\date{\today}

\begin{abstract}	
In this paper, we formulate the real-space Chern number in a supercell framework.
In this framework, the overlap matrix between two corners of the Brillouin zone (BZ) is derived from diagonalizing the real-space Hamiltonian with periodic boundary conditions.
The path-ordered product of overlap matrices around the BZ boundary forms a Wilson loop, and defines the Chern number in real space.
It is analytically shown that the real-space Chern number is quantized at integers for large enough systems and coincides with the Bott index used in the previous studies.
The formulation is greatly simplified for the former so that it makes numerical computations more efficient.
The real-space formula is used to numerically elucidate topological phases in a disordered Chern insulator.
The Chern insulator is modeled by dimensional extension of the Rice-Mele model consisting of two sublattices, and is disordered by including a random onsite potential.
As disorder strength increases, the nontrivial-to-trivial phase transition takes place for normal disorder with no sublattice polarization.
By contrast, the phase diagram is almost unaffected by polarized disorder, indicating that nontrivial topology persists against disorder.
These observations are supported by the linear conductance and the density of bulk states.
\end{abstract}

\maketitle

\section{Introduction}
\label{sec:1}

Topology hiding in quantum states of matter has attracted immense attention in recent years since it produces fundamentally new physical phenomena and potential applications in novel devices \cite{ref:1,ref:2}.
From a historical perspective, the notion of topological matter originates from the topological formulation of the integer quantum Hall effect by Thouless et al. \cite{ref:3}.
They showed that the first Chern number, a topological invariant quantized at integers, accounts for the robust quantization of Hall conductivity in two dimensions (2D).
Haldane expanded the theory of the quantum Hall effect, and introduced a model of the Chern insulator with the nonzero Chern number but without Landau levels \cite{ref:4}.
The Haldane model on a honeycomb lattice has been realized in an experiment using ultracold atoms \cite{ref:5} and in a solid-state material \cite{ref:6}.
An analogous simple model was explored by Qi, Wu and Zhang \cite{ref:7}.

The discovery of topological insulators with time-reversal symmetry led to the widespread study of topological aspects in insulators, semimetals, and superconductors \cite{ref:1, ref:2}.
In parallel to these studies, a classification scheme was established that predicts which topological invariants are definable for given systems \cite{ref:8, ref:9}.
This scheme depends only on spatial dimensions and underlying discrete symmetries including time-reversal symmetry, particle-hole symmetry, and the combination of the two known as chiral symmetry.
For instance, the quantum Hall states breaking time-reversal symmetry are classified into the unitary class A, which has a $\mathbb{Z}$ invariant in 2D, while the time-reversal-symmetric topological insulators belong to the symplectic class AII, which allows a $\mathbb{Z} _2$ invariant in 2D or 3D.

These topological invariants are usually defined in momentum space for ordered systems with translational symmetry.
A typical example is the Chern number in 2D, which can be expressed as a line integral of the Berry connection around the boundary of the Brillouin zone (BZ).
The conventional formulation in momentum space is however insufficient to elucidate topology of inhomogeneous or disordered systems that lack translational invariance.
In this sense, it is desirable to derive a generic real-space formulation that deals with ordered and disordered systems in a unified manner.
Recently, a number of approaches have been proposed in this direction, such as the switch-function formalism \cite{ref:10, ref:11, ref:12}, the scattering matrix method \cite{ref:13, ref:14}, the noncommutative index theorem \cite{ref:15, ref:16, ref:17}, structural spillage \cite{ref:18}, the Bott index \cite{ref:19, ref:20, ref:21, ref:22, ref:23, ref:24, ref:25, ref:26, ref:27, ref:28}, the single-point formula \cite{ref:29, ref:30}, and the local Chern marker \cite{ref:31, ref:32, ref:33, ref:34, ref:35, ref:36, ref:37, ref:38, ref:39}.
 
In this paper, we derive a real-space Chern number from the Wilson loop in a supercell framework.
It is analytically shown that the real-space invariant is quantized at integers for large enough systems and coincides with the Bott index used in the previous studies.
The formulation is greatly simplified for the former so that it makes numerical computations more efficient.
The real-space formula is used to numerically elucidate topological phases in a disordered Chern insulator consisting of two sublattices.
As disorder strength increases, the nontrivial-to-trivial phase transition takes place for normal disorder with no sublattice polarization.
By contrast, the phase diagram is almost unaffected by polarized disorder, indicating that nontrivial topology persists against disorder.

The paper is organized as follows.
In Sec. \ref{sec:2}, we construct a model of the Chern insulator by dimensional extension of the Rice-Mele (RM) model.
In Sec. \ref{sec:3}, we derive a theory of the real-space Chern invariant in a supercell framework.
In terms of the real-space formula, we numerically elucidate topological phases of the Chern insulator subjected to a random onsite potential in Sec. \ref{sec:4}.
Finally, Sec. \ref{sec:5} provides a summary.

\section{Rice-Mele model and Chern insulator}
\label{sec:2}

The Su-Schrieffer-Heeger (SSH) model forms a prototype for topological insulators in 1D, which consists of a bipartite lattice with two sublattice sites $A$ and $B$ in each unit cell \cite{ref:40}.
This model preserves chiral symmetry and belongs to the BDI class, which possesses a $\mathbb{Z}$ invariant known as the winding number in 1D.
The SSH model is generalized to the RM model by including a staggered sublattice potential \cite{ref:41}.
The onsite potential breaks chiral symmetry so that the RM model falls in the AI class.
The RM model is described by the lattice Hamiltonian
\begin{equation}
\label{eq:1}
\begin{split}
{H_{{\text{RM}}}} &= \sum\limits_j {\ketbra{j}{j} \otimes (v{\sigma _x} + m{\sigma _z})}  \\ 
&+ w\sum\limits_j {(\ketbra{j+1}{j} \otimes \frac{{{\sigma _x} + i{\sigma _y}}}{2} + {\text{H.c.}})}  \\ 
\end{split} ,
\end{equation}
for noninteracting spinless fermions, where $\left| j \right\rangle $ represents the basis ket at each unit cell labeled by $j \in \{ 1,2, \cdots ,{N_x}/2\} $, $v$ ($w$) denotes the intracell (intercell) hopping energy, and $m$ refers to the staggered sublattice potential.
The Pauli matrices ${\sigma _{x,y,z}}$ operate in the subspace of two sublattice sites.

A 2D lattice model is constructed by multiple RM wires, which are each parallel to the $x$ axis, equally spaced along the $y$ axis, and coupled to each other by short-range hopping interactions \cite{ref:42}.
In this study, we consider such a 2D lattice described by
\begin{equation}
\label{eq:2}
\begin{split}
H &= \sum\limits_{jl} {\ketbra{j,l}{j,l} \otimes (v{\sigma _x} + m{\sigma _z})}  \\ 
&+ w\sum\limits_{jl} {(\ketbra{j+1,l}{j,l} \otimes \frac{{{\sigma _x} + i{\sigma _y}}}{2} + {\text{H.c.}})}  \\ 
&+ t\sum\limits_{jl} {(\ketbra{j,l+1}{j,l} \otimes \frac{{{\sigma _x} + i{\sigma _z}}}{2} + {\text{H.c.}})}  \\ 
\end{split} ,
\end{equation}
where $l \in \{ 1,2, \cdots ,{N_y}\} $ refers to the wire position in the $y$ direction, and $t$ describes the interwire short-range hopping.
Applying double Fourier transformation to the 2D model in a torus geometry, the two-band Hamiltonian is formulated in momentum space as $H({\mathbf{k}}) = {\mathbf{h}}({\mathbf{k}}) \cdot \boldsymbol{\sigma} $, where ${\mathbf{k}} = ({k_x},{k_y})$ and $\boldsymbol{\sigma} = ({\sigma _x},{\sigma _y},{\sigma _z})$.
The vector ${\mathbf{h}} = ({h_x},{h_y},{h_z})$ is composed of ${h_x} = v + w \cos {k_x} + t \cos {k_y}$, ${h_y} = w \sin {k_x}$, and ${h_z} = m + t \sin {k_y}$.
The eigenequation $H({\mathbf{k}})\ket{{u_n}({\mathbf{k}})} = {\varepsilon _n}({\mathbf{k}})\ket{{u_n}({\mathbf{k}})} $ is solved to be ${\varepsilon _n} = n h$ and
\begin{equation}
\label{eq:3}
\ket{u_n} = \frac{1}{\sqrt {2h(h-n{h_z})}}{\pmqty{n({h_x}-i{h_y}) \\ h-n{h_z}}} ,
\end{equation}
where $n = \pm 1$ denotes the band index, and $h = \sqrt {h_x^2 + h_y^2 + h_z^2} $.
The second term in ${h_z}$ breaks time-reversal symmetry so that this model belongs to the D class for $m = 0$ and the A class for $m \ne 0$.
In both classes, band topology is characterized by the Chern number in 2D.

\begin{figure}
\centering
\includegraphics{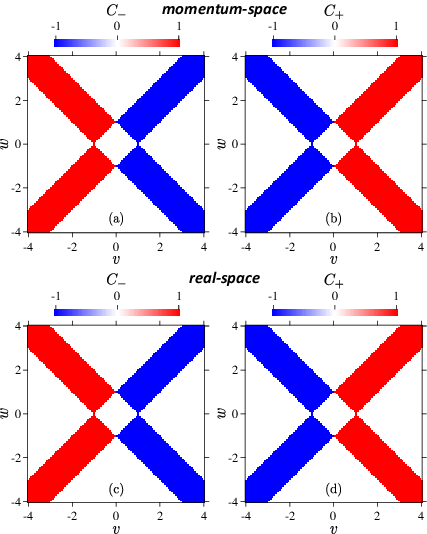}
\caption{Phase diagrams of Chern numbers ${C_ \pm }$ in the parameter space $(v,w)$. The upper two panels show the momentum-space invariants for (a) lower and (b) upper bands. The lower two panels display the real-space invariants for (c) occupied and (d) unoccupied states in the absence of disorder. The parameters used in the calculation are $(t,m,\mu ) = (1,0,0)$. The system size is $N = 40$.}
\label{fig:1}
\end{figure}

It is well known that Thouless pumping is enabled in the RM model when the system parameters are modulated periodically and slowly with time \cite{ref:42, ref:43, ref:44}.
Replacing the momentum variable ${k_y}$ with the cyclic time variable, the components of ${\mathbf{h}}$ given above coincide with those for a pumping sequence.
This relevance implies that the 2D model considered here forms a Chern insulator with the nonzero Chern number.
The momentum-space Chern number can be expressed as $C_n = \tfrac{1}{2\pi }{\smallint _{\text{BZ}}}{d^2}k{({\nabla _{\mathbf{k}}} \times {{\mathbf{A}}_n})_z}$, where ${\mathbf{A}}_n = i\braket{u_n}{{\nabla _{\mathbf{k}}} {u_n}} $ is the Berry connection.
Figures \ref{fig:1} (a) and \ref{fig:1} (b) show ${C_ \pm }$ derived from Eq. (\ref{eq:3}) in the parameter space $(v,w)$ with the assumption of $(t,m) = (1,0)$.
Note that ${C_n}$ obeys the sum rule ${\sum _n}{C_n} = 0$.
The phase-transition point is identical to the gap-closing point, at which ${(v \pm t)^2} = {w^2}$ for $m = 0$.
The Chern number becomes a nonzero integer ${C_n} = \mathrm{sgn} (nv)$ in the region where ${(v \mp t)^2} < {w^2} < {(v \pm t)^2}$.
The nontrivial phases also emerge for $m \ne 0$.
For instance, the nonzero Chern number ${C_n} = \mathrm{sgn} (nv)$ is found in the region where $0 < {t^2} - {m^2} < 4{v^2}$ for ${v^2} = {w^2}$ (not shown).
The present model is equivalent to the previous one introduced by Qi, Wu and Zhang.
The equivalence is shown by the unitary rotation $R = {e^{i\pi {\sigma _y}/4}}{e^{i\pi {\sigma _x}/4}}$.
The transformed Hamiltonian $\tilde H = RH{R^\dag } = {\tilde{\mathbf{h}}} \cdot \boldsymbol{\sigma} $ is described by ${\tilde h_x} = t \sin {k_x}$, ${\tilde h_y} = t \sin {k_y}$, and ${\tilde h_z} = v + t (\cos {k_x} + \cos {k_y})$, where $w = t$ and $m = 0$ are assumed for simplicity.
The Hamiltonian $\tilde H$ is similar to that given in the literature \cite{ref:7, ref:42}.

\begin{figure}
\centering
\includegraphics{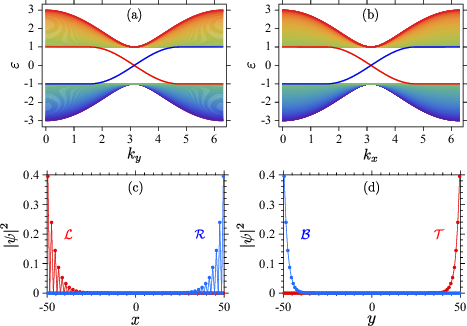}
\caption{The upper two panels show energy eigenvalues of (a) $H({k_y})$ and (b) $H({k_x})$ for cylindrical models with open ends. As shown in each figure, two chiral edge modes traverse the gap. The lower two panels display the probability densities of edge modes at momenta (c) ${k_y} = 4\pi /7$ and (d) ${k_x} = 4\pi /7$. In each panel, dots and solid lines represent numerical and analytical results, respectively. The parameters used in the calculation are $(v,w,t,m) = (1,1,1,0)$. The cylinder lengths are ${N_x} = 100$ for (a) and (c), and ${N_y} = 100$ for (b) and (d).}
\label{fig:2}
\end{figure}

Assuming the 2D model in a cylindrical geometry, a 1D lattice Hamiltonian indexed by ${k_x}$ or ${k_y}$ is obtained after performing partial Fourier transformation.
For example, the ${k_y}$-dependent Hamiltonian is written as
\begin{equation}
\label{eq:4}
\begin{split}
H({k_y}) &= \sum\limits_j {\ketbra{j}{j} \otimes ({v_y}{\sigma _x} + {h_z}{\sigma _z})}  \\
&+ w\sum\limits_j {(\ketbra{j+1}{j} \otimes \frac{{{\sigma _x} + i{\sigma _y}}}{2} + {\text{H.c.}})}  \\ 
\end{split} ,	
\end{equation}
where ${v_y} = v + t \cos {k_y}$.
For a long enough cylinder with open ends, the eigenfunctions are analytically derived for two edge modes to be
\begin{equation}
\label{eq:5}
\ket{\mathcal{L}} \propto \sum\limits_j {\ket{j,A} {{( - \frac{{{v_y}}}{w})}^{j - 1}}} ,
\end{equation}
\begin{equation}
\label{eq:6}
\ket{\mathcal{R}} \propto \sum\limits_j {\ket{j,B} {{( - \frac{{{v_y}}}{w})}^{{N_x}/2 - j}}} .
\end{equation}
These solutions fulfill the eigenequations $H({k_y})\ket{\mathcal{L}} = {h_z}\ket{\mathcal{L}} $ and $H({k_y})\ket{\mathcal{R}} = - {h_z}\ket{\mathcal{R}} $.
Note that both edge modes are ${\sigma _z}$-polarized.
Analogous results are obtained for a single RM wire \cite{ref:44, ref:45}.
The ${k_x}$-dependent Hamiltonian is given by
\begin{equation}
\label{eq:7}
\begin{split}
H({k_x}) &= \sum\limits_l {\ketbra{l}{l} \otimes ({v_x}{\sigma _x} + {h_y}{\sigma _y} + m{\sigma _z})}  \\ 
&+ t\sum\limits_l {(\ketbra{l+1}{l} \otimes \frac{{{\sigma _x} + i{\sigma _z}}}{2} + {\text{H.c.}})}  \\ 
\end{split} ,
\end{equation}
where ${v_x} = v + w \cos {k_x}$.
In this case, one finds the eigenfunctions for two edge modes, expressed as
\begin{equation}
\label{eq:8}
\ket{\mathcal{T}} \propto \sum\limits_l {\ket{l,A'} {{( - \frac{{u_x^*}}{t})}^{{N_y} - l}}} ,
\end{equation}
\begin{equation}
\label{eq:9}
\ket{\mathcal{B}} \propto \sum\limits_l {\ket{l,B'} {{( - \frac{{{u_x}}}{t})}^{l - 1}}} ,
\end{equation}
where ${u_x} = {v_x} + i m$, $\ket{A'} = (\ket{A} + i\ket{B})/\sqrt 2 $, and $\ket{B'} = (\ket{B} + i\ket{A} )/\sqrt 2 $.
Note that ${\sigma _y}\ket{A'} = \ket{A'} $ and ${\sigma _y}\ket{B'} = - \ket{B'} $.
Thus, both edge modes are ${\sigma _y}$-polarized.
It is easy to show that $H({k_x})\ket{\mathcal{T}} = {h_y}\ket{\mathcal{T}} $ and $H({k_x})\ket{\mathcal{B}} = - {h_y}\ket{\mathcal{B}} $.
As seen in Fig. \ref{fig:2}, these analytical results agree with the numerical results derived for $(v,w,t,m) = (1,1,1,0)$.
The existence conditions are $\abs{{v_y}/w} < 1$ for $\ket{\mathcal{L}} $ and $\ket{\mathcal{R}} $, and $\abs{{u_x}/t} < 1$ for $\ket{\mathcal{T}} $ and $\ket{\mathcal{B}} $.
These two conditions are reduced to ${k_{x,y}} \in (\tfrac{\pi }{2},\tfrac{{3\pi }}{2})$ for the parameters assumed in the calculation.
In this range, sinusoidal dispersion curves traverse the band gap for the chiral edge modes.

\section{Real-space Chern invariant}
\label{sec:3}

The Chern number is defined as a line or surface integral in momentum space.
The conventional formulation in momentum space is however insufficient to elucidate topology of inhomogeneous or disordered systems lacking translational invariance.
In this section, we generalize the theoretical treatment to derive a real-space formula, which can deal with ordered and disordered systems on an equal footing.

In a supercell framework, we usually assume a fictitious periodic system that consists of a large unit cell.
Treating a disordered system as a supercell, this approach introduces the momentum-space eigenequation $H({\mathbf{k}})\ket{{u_n}({\mathbf{k}})} = {\varepsilon _n}({\mathbf{k}})\ket{{u_n}({\mathbf{k}})} $ to the disordered system.
Here, we consider a supercell of size ${N_x} \times {N_y}$.
Then, the primitive translation vector of the reciprocal lattice is given by ${{\mathbf{b}}_\nu } = \tfrac{{2\pi }}{{{N_\nu }}}{{\mathbf{e}}_\nu }$, where $\nu = x,y$, and ${{\mathbf{e}}_\nu }$ is the unit vector in the $\nu $ direction.
The position operator is expressed as ${\mathbf{r}} = i{\nabla_{\mathbf{k}}}$ in momentum space so that $\ket{{u_n}({{\mathbf{b}}_\nu })} = {e^{ - i{{\mathbf{b}}_\nu } \cdot {\mathbf{r}}}}\ket{{u_n}(0)} $.
Since the momentum translation operator ${e^{ - i{{\mathbf{b}}_\nu } \cdot {\mathbf{r}}}}$ is unitary, the above translation ensures the periodicity ${\varepsilon _n}({{\mathbf{b}}_\nu }) = {\varepsilon _n}(0)$ in the BZ.
Note that $H(0)$ is identical to the real-space Hamiltonian $H$ of the supercell with periodic boundary conditions.
Thus, $\left| {{u_n}(0)} \right\rangle $ is derived by numerically solving the real-space eigenequation $H\ket{{u_n}} = {\varepsilon _n}\ket{{u_n}} $.

In what follows, we restrict our attention to either occupied or unoccupied states.
The subspace formed by these states is distinguished by, e.g., $s = - 1$ for ${\varepsilon _n} < \mu $ or $s = 1$ for ${\varepsilon _n} > \mu $, respectively, where $\mu $ denotes the chemical potential.
The theoretical treatments are equivalent for these two subspaces.
For this reason, the index $s$ is suppressed in the following arguments to avoid unnecessary complications.
In the supercell framework, four corners of the BZ are at ${{\mathbf{k}}_0} = (0,0)$, ${{\mathbf{k}}_1} = ({b_x},0)$, ${{\mathbf{k}}_2} = ({b_x},{b_y})$, and ${{\mathbf{k}}_3} = (0,{b_y})$.
First, we introduce the overlap matrix between two corners such that
\begin{equation}
\label{eq:10}
q_{nn'}^{\alpha \beta } = \braket{{u_n}({\mathbf{k}}_\alpha )}{{u_{n'}}({\mathbf{k}}_\beta )}
 = \bra{u_n} {e^{i({{\mathbf{k}}_\alpha } - {{\mathbf{k}}_\beta }) \cdot {\mathbf{r}}}} \ket{u_{n'}} .
\end{equation}
It is more convenient to use matrix notation, in which Eq. (\ref{eq:10}) is expressed as 
\begin{equation}
\label{eq:11}
{q^{\alpha \beta }} = {U^\dag }{e^{i({{\mathbf{k}}_\alpha } - {{\mathbf{k}}_\beta }) \cdot {\mathbf{r}}}}U ,
\end{equation}
where ${U_{{\mathbf{r}}n}} = \braket {\mathbf{r}}{{u_n}} $ is the rectangular matrix composed of occupied (or unoccupied) eigenstates, and ${\mathbf{r}} = (x,y)$ denotes the site position.
Second, we construct the path-ordered product of overlap matrices
\begin{equation}
\label{eq:12}
Q = {q^{03}}{q^{32}}{q^{21}}{q^{10}} .
\end{equation}
Note that ${q^{\alpha \beta }} = {e^{i({{\mathbf{k}}_\alpha } - {{\mathbf{k}}_\beta }) \cdot {\mathbf{A}}({{\mathbf{k}}_\beta })}}$ and hence $Q = \mathcal{P}{e^{i\oint_{\partial {\text{BZ}}} {d{\mathbf{k}} \cdot {\mathbf{A}}({\mathbf{k}})} }}$ in the large supercell limit, where ${{\mathbf{A}}_{nn'}}({\mathbf{k}}) = i\braket {{u_n}({\mathbf{k}})}{{\nabla_{\mathbf{k}}} {u_{n'}}({\mathbf{k}})} $ is the Berry connection matrix, $\mathcal{P}$ refers to the path-ordering operator, and $\partial {\text{BZ}}$ denotes the BZ boundary.
Thus, the Chern number is definable as
\begin{equation}
\label{eq:13}
C = \frac{1}{{2\pi i}} \Tr \ln Q = \frac{1}{{2\pi i}} \ln \det Q .
\end{equation}
Since ${q^{\alpha \beta }}$ and $Q$ are unitary in the limit of ${N_\nu } \to \infty $, we finally arrive at
\begin{equation}
\label{eq:14}
C = \frac{1}{{2\pi }}\sum\limits_p {\arg {\lambda _p}} ,
\end{equation}
where ${\lambda _p}$ is the eigenvalue of $Q$.

It is easy to show that $C$ is quantized to be an integer for a large enough supercell.
Note that ${q^{32}} = {({q^{10}})^\dag }$ and ${q^{03}} = {({q^{21}})^\dag }$ by definition.
In terms of these correspondences, $\det Q = 1$ in the ${N_\nu } \to \infty $ limit, where ${q^{\alpha \beta }}$ is unitary.
Hence, ${\sum _p} \arg {\lambda _p} = 2\pi \ell $ so that $C = \ell $, where $\ell \in \mathbb{Z}$.
This brief proof is essentially the same as that for the Bott index of a pair of unitary matrices, which is always an integer \cite{ref:27}.

As described above, the real-space Chern number is derived from a Wilson loop formed by the path-ordered product of overlap matrices around the BZ boundary.
It may be appropriate here to compare this method to the conventional ones, such as the single-point Chern number and the Bott index.

The single-point Chern number \cite{ref:29, ref:30, ref:36} is formulated as
\begin{equation}
\label{eq:15}
C_{\text{asym}} = - \frac{1}{\pi } \Im\sum\limits_n \braket{{{\tilde u}_n}({{\mathbf{b}}_x})}{{{\tilde u}_n}({{\mathbf{b}}_y})}
\end{equation}
or
\begin{equation}
\label{eq:16}
C_{\text{sym}} = - \frac{1}{4\pi } \Im\sum\limits_{nrr'} rr' \braket{{{\tilde u}_n}(r {{\mathbf{b}}_x})}{{{\tilde u}_n}( r' {{\mathbf{b}}_y})} ,
\end{equation}
where $\ket{{{\tilde u}_n}({{\mathbf{b}}_\nu })} = \sum _{n'} \ket{{u_{n'}}({{\mathbf{b}}_\nu })} q_{n'n}^{-1} ({{\mathbf{b}}_\nu }) $
, ${q_{nn'}}({{\mathbf{b}}_\nu }) = \braket{u_n}{u_{n'} ({{\mathbf{b}}_\nu }) } $, and $r = \pm 1$.
These two formulas consist of a single overlap matrix and a linear combination of overlap matrices, respectively.
Thus, the single-point Chern number essentially differs from the real-space Chern number defined by the path-ordered product of overlap matrices.

The Bott index is defined for a pair of almost unitary matrices in the form \cite{ref:19, ref:20, ref:21, ref:22, ref:23, ref:24, ref:25, ref:26, ref:27, ref:28}
\begin{equation}
\label{eq:17}
{I_{\text{B}}} = \frac{1}{{2\pi i}} \Tr \ln {V_y}{V_x}V_y^\dag V_x^\dag .
\end{equation}
Originally, ${V_\nu } = P{e^{i{{\mathbf{b}}_\nu } \cdot {\mathbf{r}}}}P$ is assumed, where $P = U{U^\dag }$ is the projector onto the subspace of occupied (or unoccupied) eigenstates.
Then, ${I_{\text{B}}}$ is rewritten as
\begin{equation}
\label{eq:18}
{I_{\text{B}}} = \frac{1}{{2\pi i}} \ln \det UQ{U^\dag } ,
\end{equation}
so that ${I_{\text{B}}} \ne C$.
Instead, the recent studies adopt ${V_\nu } = P{e^{i{{\mathbf{b}}_\nu } \cdot {\mathbf{r}}}}P + (1 - P)$ to improve the numerical stability of the algorithm.
In this case, we find
\begin{equation}
\label{eq:19}
{I_{\text{B}}} = \frac{1}{{2\pi i}} \ln \det [1 + U(Q - 1){U^\dag }] .
\end{equation}
Subsequently, the Sylvester's determinant identity $\det (1 + AB) = \det (1 + BA)$ leads us to
\begin{equation}
\label{eq:20}
{I_{\text{B}}} = \frac{1}{{2\pi i}} \ln \det Q ,
\end{equation}
so that ${I_{\text{B}}} = C$.
Thus, the stabilized version of the Bott index is reduced to the real-space Chern number derived here.
It is worth noting that the latter makes numerical computations more efficient since no projection is involved in its formulation.

Toniolo showed that the Bott index is equivalent to the Chern number of the Fermi projection $P$, written as
\begin{equation}
\label{eq:21}
\operatorname{Ch} (P) = 4\pi \Im \Tr_{\text{u.a.}} P[x,P][y,P] ,
\end{equation}
where $\Tr_{\text{u.a.}} $ denotes the trace per unit area \cite{ref:27}.
Thus, the present study ultimately shows the equality of the three formulations, namely, the Wilson-loop formulation of the Chern number in real space, the Bott index of a pair of almost unitary matrices, and the Chern number of the projection.

\begin{figure}
\centering
\includegraphics{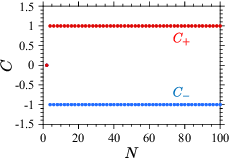}
\caption{Real-space Chern numbers ${C_ \pm }$ as a function of system size $N$ in the absence of disorder. The parameters used in the calculation are $(v,w,t,m,\mu ) = (1,1,1,0,0)$.}
\label{fig:3}
\end{figure}

The real-space Chern number is compared to that in momentum-space in Fig. \ref{fig:1}.
The real-space invariant is numerically derived for a periodic ordered system of size $N = {N_x} = {N_y} = 40$.
As shown in the figure, the two types of topological invariants are identically quantized.
Figure \ref{fig:3} explains the size dependence.
It is found from the figure that Eq. (\ref{eq:14}) correctly evaluates the Chern number for moderately large systems.
The numerical accuracy is typically on the order of ${10^{ - 14}}$.
Note that the bulk invariant is derived for $N \ge 4$, indicating that it is actually unnecessary to approach the thermodynamic limit $N \to \infty $.
This observation does not contradict Eq. (\ref{eq:14}).
Since $\abs{\arg \lambda _p} \le \pi $, one expects $\abs{C} \le {N^2}/4$ at half-filling. Thus, $\abs{C} = 1$ is reproducible for $N \ge 2$ in the present method.

\begin{figure}
\centering
\includegraphics{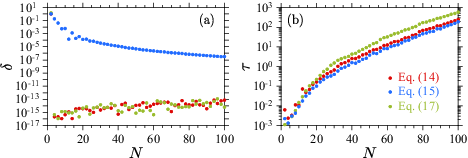}
\caption{Numerical evaluations of Eqs. (\ref{eq:14}), (\ref{eq:15}) and (\ref{eq:17}). Two panels show (a) the absolute error $\delta $ and (b) the execution time $\tau $ of computation as a function of system size $N$ in the absence of disorder. The parameters used in the calculation are $(v,w,t,m,\mu ) = (1,1,1,0,0)$.}
\label{fig:4}
\end{figure}

We have also numerically evaluated Eqs. (\ref{eq:14}), (\ref{eq:15}) and (\ref{eq:17}).
The numerical results are summarized in Fig. \ref{fig:4}.
In general, Eq. (\ref{eq:15}) produces erroneous results in comparison with Eq. (\ref{eq:14}).
As shown in Fig. \ref{fig:4} (a), there is a considerable difference in the absolute error $\delta $ (i.e., the absolute value of the difference between computed and theoretical values).
For the present model, $\delta $ is similar for Eqs. (\ref{eq:15}) and (\ref{eq:16}).
More quantitatively, $\delta $ behaves as $29 \times {N^{ - 4}}$ for the two formulas in the large $N$ limit.
This result is different from that for the Haldane model on a honeycomb lattice \cite{ref:36}, implying that the decay characteristics depend on the electronic structures of the model assumed.
On the other hand, Eq. (\ref{eq:17}) involves many matrix products including the projector $P = U{U^\dag }$ so that its computation tends to be time-consuming in comparison with Eq. (\ref{eq:14}).
As shown in Fig. \ref{fig:4} (b), there is an appreciable difference in the execution time $\tau $ of computation for large systems.

\section{Disordered Chern insulators}
\label{sec:4}

Next, we proceed to numerical calculations of disordered Chern insulators.
Considering the sublattice degrees of freedom $\sigma \in \{ A,B\} $, a random onsite potential is expressed generally as \cite{ref:45}
\begin{equation}
\label{eq:22}
{H_{{\text{rand}}}} = \sum\limits_{jl\sigma } {\ket{j,l,\sigma } {u_{jl\sigma }} \bra{j,l,\sigma }} .
\end{equation}
We regard ${u_{jl\sigma }} \in [ - {W_\sigma },{W_\sigma }]$ as a random variable following the uniform distribution.
We assume $W = {W_A} = {W_B}$ for normal disorder, and either ${W_A} = 0$ or ${W_B} = 0$ for polarized disorder.
As expected, similar results are derived for ${W_A} = 0$ and ${W_B} = 0$.
The disordered Chern insulator is described by adding ${H_{{\text{rand}}}}$ to Eq. (\ref{eq:2}) as an extra term.
The parameters used in the calculations are typically $(v,w,t,m,\mu ) = (1,1,1,0,0)$.
The system size is chosen to be $N = 40 - 100$ unless stated otherwise.

\begin{figure}
\centering
\includegraphics{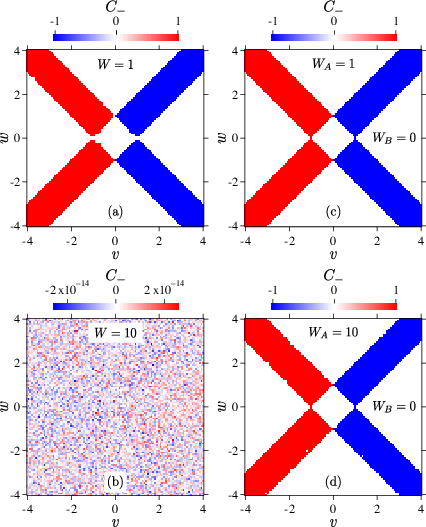}
\caption{Real-space Chern number ${C_ - }$ in the parameter space $(v,w)$ in the presence of disorder. The left two panels show ${C_ - }$ for normal disorder at (a) $W = 1$ and (b) $W = 10$. The right two panels display ${C_ - }$ for polarized disorder at (c) ${W_A} = 1$ and (d) ${W_A} = 10$. The parameters used in the calculation are $(t,m,\mu ) = (1,0,0)$. The system size is $N = 40$.}
\label{fig:5}
\end{figure}

\begin{figure}
\centering
\includegraphics{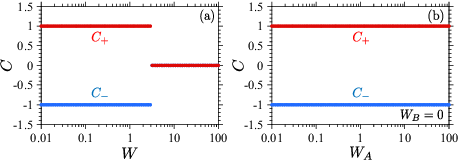}
\caption{Real-space Chern numbers ${C_ \pm }$ as a function of disorder strength. Two panels show the numerical results for (a) normal disorder and (b) polarized disorder, where the disorder strengths are denoted by $W$ and ${W_A}$, respectively. The parameters used in the calculation are $(v,w,t,m,\mu ) = (1,1,1,0,0)$. The system size is $N = 100$.}
\label{fig:6}
\end{figure}

Figure \ref{fig:5} shows how the real-space Chern number ${C_ - }$ for occupied states varies in $(v,w)$ space at various disorder strengths.
Each data point corresponds to a single realization of the random potential.
As seen in Fig. \ref{fig:5} (a), except for a small discrepancy around the phase boundaries, ${C_ - }$ remains identical to that for ordered systems as long as normal disorder is weak enough.
On the other hand, ${C_ - }$ vanishes in the entire parameter space as $W$ exceeds a critical value, as shown in Fig. \ref{fig:5} (b).
Thus, normal disorder induces the nontrivial-to-trivial phase transition.
By contrast, the phase diagram is almost unaffected by polarized disorder, as shown in Figs. \ref{fig:5} (c) and \ref{fig:5} (d).
Figure \ref{fig:6} shows ${C_ \pm }$ as a function of disorder strength.
The phase transition takes place at $W \simeq 3$ for normal disorder, while ${C_ \pm }$ is invariable up to ${W_A} = 100$ for polarized disorder.
Irrespective of the type of disorder, disorder-strength dependence is unaffected by the $m$ term as long as ordered systems remain nontrivial (not shown).
In view of these results, we can say that the disorder-induced phase transition is absent for polarized disorder.

\begin{figure}
\centering
\includegraphics{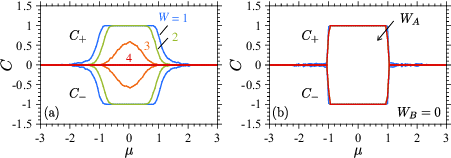}
\caption{Disorder-averaged ${C_ \pm }$ as a function of the chemical potential $\mu $ for (a) normal disorder and (b) polarized disorder. The disorder strength is varied as $W = 1,2,3,4$ in (a), and ${W_A} = 1,10,100$ in (b). The parameters used in the calculation are $(v,w,t,m) = (1,1,1,0)$. The system size is $N = 40$.}
\label{fig:7}
\end{figure}

In the absence of disorder, two nontrivial bands with opposite Chern numbers ${C_ \pm } = \pm 1$ are separated by a finite spectral gap.
As disorder strength increases, the spectral gap is reduced.
Eventually, the two bands merge together, and all the states are localized.
Then, the ${C_ \pm }$ mix and totally vanish.
This corresponds to the well-known levitation-annihilation process that explains the nontrivial-to-trivial phase transition due to disorder.
Related spectral information is derived from varying the chemical potential $\mu $.
Figure \ref{fig:7} shows the disorder-averaged ${C_ \pm }$ as a function of $\mu $.
In this calculation, disorder averaging is performed over ${10^3}$ random configurations.
As seen in the figure, there is an interval where ${C_ \pm } = \pm 1$ are maintained.
In this interval, statistical fluctuations are negligibly small (not shown).
For brevity, we refer to it as a nontrivial interval.
As shown in Fig. \ref{fig:7} (a), the nontrivial interval shrinks as $W$ increases for normal disorder, and finally disappears as $W$ becomes large enough.
The nontrivial interval correlates with the finite gap.
Thus, the behavior shown in Fig. \ref{fig:7} (a) supports the above argument.
On the other hand, the nontrivial interval persists irrespective of ${W_A}$ for polarized disorder, as shown in Fig. \ref{fig:7} (b).
In this case, the nontrivial interval discontinuously connects to the trivial ones where ${C_ \pm } = 0$.
These features can be accounted for by nontrivial band-edge states and their persistence against disorder.

The physical insights derived from ${C_ \pm }$ in the presence of disorder are corroborated by the two-terminal linear conductance $G$ and the density of bulk states per site $D$.
Recursive Green's functions are used to calculate $G$ and $D$ \cite{ref:45, ref:46, ref:47, ref:48}.
In the calculation of $G$, semi-infinite ideal leads (ordered Chern insulator) are attached at two longitudinal ends of a strip running along $y$.
In this geometry, open boundary conditions are imposed at two lateral edges $x = \pm {N_x}/2$, where edge states are formed.
In the calculation of $D$, a cylindrical geometry is assumed, where edge states are eliminated because of periodic boundary conditions.
The length of the cylinder is chosen to be ${N_y} = {10^5}$ for numerical convergence.

\begin{figure}
\centering
\includegraphics{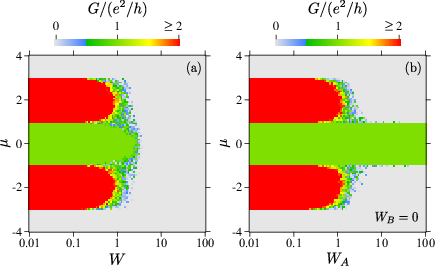}
\caption{Linear conductance $G$ of a strip model of size $N = 100$ as a function of disorder strength and chemical potential $\mu $. Two panels show the numerical results for (a) normal disorder and (b) polarized disorder, where the disorder strengths are denoted by $W$ and ${W_A}$, respectively. The parameters used in the calculation are $(v,w,t,m) = (1,1,1,0)$.}
\label{fig:8}
\end{figure}

\begin{figure}
\centering
\includegraphics{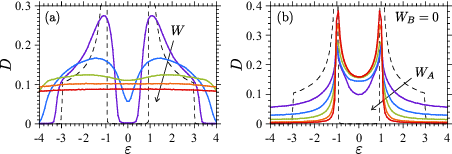}
\caption{Density of states $D$ in a cylindrical model of circumference ${N_x} = 100$ and length ${N_y} = {10^5}$ as a function of energy $\varepsilon $. Two panels show the numerical results for (a) normal disorder and (b) polarized disorder. The disorder strength is varied as $W = 1,2,3,4,5$ in (a), and ${W_A} = 5,10,20,50,100$ in (b). The parameters used in the calculation are $(v,w,t,m) = (1,1,1,0)$. A thin dashed line in each figure shows $D$ in the absence of disorder as a reference.}
\label{fig:9}
\end{figure}

Figure \ref{fig:8} (a) shows the linear conductance $G$ as a function of $W$ and $\mu $ for normal disorder.
When $W$ is small enough, $G$ remains quantized to be ${e^2}/h$ due to the 1D conduction through edge states in the gap.
The 2D conduction via bulk bands out of the gap is greatly suppressed for $W \gtrsim 1$, whereas edge conduction survives against disorder up to $W \simeq 3$.
As $W$ further increases, $G$ totally vanishes, indicating that all the states are localized.
On the other hand, edge conduction is unaffected by polarized disorder, as shown in Fig. \ref{fig:8} (b).
In terms of the bulk-edge correspondence, this implies that there exist nontrivial bulk states, and they are insusceptible to disorder.
Figure \ref{fig:9} (a) shows the density of bulk states $D$ in the presence of normal disorder.
As seen is the figure, the two bands merge into a single broad band for strong disorder.
Such features are not observed for polarized disorder.
In this case, two sharp peaks remain at the band edges even for strong disorder, as shown in Fig. \ref{fig:9} (b).
Thus, it is reasonable to consider that persistent band-edge states suppress the mixing and annihilation of ${C_ \pm }$ for polarized disorder.

Note that $\bra{\mathcal{L}} {H_{\text{rand}}} \ket{\mathcal{L}} = 0$ for ${W_A} = 0$, and similarly $\bra{\mathcal{R}} {H_{\text{rand}}} \ket{\mathcal{R}} = 0$ for ${W_B} = 0$.
This implies that one of these two edge modes tends to survive in the presence of polarized disorder.
In view of the bulk-edge correspondence, the persistence of edge mode should be accompanied by the nontrivial bulk topology, which is identified by the nonzero Chern number.
In this way, we can understand our findings phenomenologically, although further studies are needed for their quantitative explanation.

\section{Summary}
\label{sec:5}

We have formulated the real-space Chern number in a supercell framework.
In this framework, the overlap matrix between two corners of the BZ is derived from diagonalizing the real-space Hamiltonian under periodic boundary conditions.
The Chern number in real space is defined by the Wilson loop formed by the path-ordered product of overlap matrices around the BZ boundary.
For large enough systems, it can be analytically shown that the real-space Chern number is quantized at integers.
The Chern number formulated in this way coincides with the Bott index used in previous studies, but is greatly simplified and makes numerical computations more efficient.

The real-space formula was employed to numerically elucidate topological phases in a disordered Chern insulator.
The Chern insulator was modeled by dimensional extension of the RM model, and was disordered by including a random onsite potential.
Increasing the disorder strength produces a nontrivial-to-trivial phase transition for normal disorder.
In contrast, polarized disorder leaves the phase diagram almost unaffected.
This anomaly is accounted for by nontrivial band-edge states that persist against disorder.
The linear conductance and the density of bulk states further corroborate these findings.

\bibliography{ref}
 
\end{document}